\documentclass[journal]{vgtc}                

\ifpdf
  \pdfoutput=1\relax                   
  \usepackage{graphicx}                
  \DeclareGraphicsExtensions{.pdf,.png,.jpg,.jpeg} 
\else
  \usepackage{graphicx}                
  \DeclareGraphicsExtensions{.eps}     
\fi%

\graphicspath{{figures/}{pictures/}{images/}{./}} 

\usepackage{microtype}                 
\PassOptionsToPackage{warn}{textcomp}  
\usepackage{textcomp}                  
\usepackage{mathptmx}                  
\usepackage{times}                     
\usepackage{cite}                      
\usepackage{tabu}                      
\usepackage{booktabs}                  

\usepackage{lipsum}

\usepackage{enumitem}

\usepackage{enumitem}
\usepackage{amsmath} 
\usepackage{amssymb}  
\usepackage{fixmath}
\renewcommand{\vec}[1]{\mathbold{#1}}
\newcommand{\mat}[1]{\mathbold{#1}}
\usepackage[ruled,vlined]{algorithm2e}

\usepackage{comment}
\usepackage{subfigure}

\usepackage{fontawesome}

\newcommand{\aar}[1]{\textcolor{black}{#1}}
\newcommand{\sk}[1]{\textcolor{black}{#1}}
\newcommand{\ct}[1]{\textcolor{black}{#1}}

\newcommand{\euan}[1]{\textcolor{black}{#1}}

\onlineid{0}

\vgtccategory{Research}

\vgtcinsertpkg

\title{Propagating Visual Designs to Numerous Plots and Dashboards}

\author{%
Saiful Khan$^1$\thanks{e-mail: saiful.khan@eng.ox.ac.uk},\hspace{4mm}
Phong H. Nguyen$^2$\thanks{e-mail: {phong.nguyen@redsift.io}},\hspace{4mm}
Alfie Abdul-Rahman$^3$\thanks{e-mail: alfie.abdulrahman@kcl.ac.uk},\hspace{4mm}
Benjamin Bach$^4$\thanks{e-mail: bbach@ed.ac.uk},\hspace{4mm}
Min Chen$^1$\thanks{e-mail: min.chen@oerc.ox.ac.uk},\\
Euan Freeman$^5$\thanks{e-mail: euan.freeman@glasgow.ac.uk},\hspace{4mm}
Cagatay Turkay$^6$\thanks{e-mail: cagatay.turkay@warwick.ac.uk \vspace{-4mm}}\\
\scriptsize \centering
$^1$ University of Oxford,\hspace{2mm}
$^2$ Redsift Ltd.,\hspace{2mm}
$^3$ King's College London,\hspace{2mm}
$^4$ Edinburgh University,\hspace{2mm}
$^5$ University of Glasgow,\hspace{2mm}
$^6$ University of Warwick}

\teaser{
  \centering
  \includegraphics[width=0.9\linewidth]{figures/Teaser-Final.pdf}
  \caption{Our novel propagation workflow makes it easy to propagate visual designs to numerous datasets. \euan{Reference} visualizations are created for \euan{data streams, which are associated with} several keywords in our ontology. \euan{A search and activate process is used to propagate the reference visualisation to other appropriate data streams.} (1)~Ontology keywords are used to construct a query \euan{in our search UI} for suitable data stream combinations. (2)~Search results consist of ranked data stream combinations that match query parameters, although some results may not be suitable for propagation. (3)~A quality assurance step \euan{carried out by an expert} ensures the visual design is only propagated to suitable data, resulting in new visualizations that are immediately deployed \euan{as web pages}.}
  \label{fig:teaser}
}

\abstract{In the process of developing an infrastructure for providing visualization and visual analytics (VIS) tools to epidemiologists and modeling scientists, we encountered a technical challenge for applying a number of visual designs to numerous datasets rapidly and reliably with limited development resources. In this paper, we present a technical solution to address this challenge. Operationally, we separate the tasks of data management, visual designs, and plots and dashboard deployment in order to streamline the development workflow. Technically, we utilize: an ontology to bring datasets, visual designs, and deployable plots and dashboards under the same management framework; multi-criteria search and ranking algorithms for discovering potential datasets that match a visual design; and a purposely-design user interface for propagating each visual design to appropriate datasets (often in tens and hundreds) and quality-assuring the propagation before the deployment. This technical solution has been used in the development of the RAMPVIS infrastructure for supporting a consortium of epidemiologists and modeling scientists through visualization.}


\CCScatlist{
  \CCScatTwelve{Human-centered computing}{Visu\-al\-iza\-tion}{Visu\-al\-iza\-tion techniques}{Treemaps};
  \CCScatTwelve{Human-centered computing}{Visu\-al\-iza\-tion}{Visualization design and evaluation methods}{}
}

\renewcommand{\lipsum}[1]{}

\begin{document}


\firstsection{Introduction}
\maketitle


RAMPVIS \cite{rampvis:2020:arXiv} is a group of volunteers specialized in Visualization and Visual Analytics (VIS), who answered a call to support the modeling scientists and epidemiologists in the Scottish COVID-19 Response Consortium (SCRC). One major challenge identified at the beginning (May 2020) was that there was a huge amount of data that epidemiologists and modeling scientists in the SCRC needed to access rapidly but could only do so via data files in a variety of inconsistent formats, requiring time-consuming processing. In most cases, they had to create simple plots using spreadsheet facilities, as they lacked the expertise and tools to create visual designs tailored to their needs. Meanwhile, another group of volunteers in the SCRC started developing a data infrastructure for hosting captured data, as well as computational data resulting from model testing, uncertainty analysis, and model predication. A subset of VIS volunteers formed a \emph{generic support team} focusing on providing a VIS infrastructure coupled with the SCRC data infrastructure, while other VIS volunteers gathered detailed domain-specific requirements for analytical, model-developmental, and disseminative visualization, and developed techniques for supporting these visualization needs. 

Although the generic support team managed to bring together a large cohort of UK-based VIS volunteers who could contribute to the design and engineering of a VIS system for visualizing thousands of datasets, the total programming resources in the team was less than one full-time person. The team thus streamlined their development effort by componentizing different tasks in the workflow, from obtaining data streams from the SCRC data infrastructure, to making visualization facilities available to the end users. By July 2020, the team produced simple plots and dashboards, and could semi-automatically propagate them to numerous data streams of similar data types (e.g., propagating a plot across data from all regions of a country). In this context, propagation means binding a visualization function to numerous data streams to deploy new interactive visualizations (as in \autoref{fig:teaser}). This helped shape the VIS infrastructure through rapid development with minimal cost.

However, since then, many more datasets have become available and more complex plots and dashboards have been added to the VIS infrastructure. The process of propagating these to new data streams becomes more complex, requiring more advanced support and quality assurance. 
\aar{While existing systems (e.g., Tableau or Power BI) provide powerful user interfaces for creating plots and dashboards, they do not provide a propagation mechanism for transforming the design of a plot or dashboard with multiple data sources to many hundreds of plots or dashboards with similar but less-well-defined data sources in an efficient, scalable, quality-assured manner.}

For example, as illustrated in \autoref{fig:teaser}, a stack bar chart was developed by a VIS volunteer for juxtaposing six time series representing the fatalities in different location types (e.g., care home, hospital) in a geographic region. As over 300 regions in the UK have similar time series data, another VIS volunteer, a dedicated \textit{infrastructure manager}, used the existing stack bar chart as a template, searched for all possible sets of six time series that match with the template, checked for errors in the search results, and finally activated the propagation individually or in groups. The result was several new interactive visualizations covering all regions, created, deployed, and linked to other visualizations with minimal development time and cost. In this paper, we present the development of the VIS infrastructure that enables this cost-effective process for propagating visual designs to numerous plots and dashboards.

Our main contributions include:
\begin{itemize}[noitemsep,nolistsep]
    \item A novel design of an \textbf{ontology-based infrastructure} for enabling search for matching data sets;
    \item A \textbf{streamlined workflow} that helps deploy the limited programming resources cost-effectively;
    \item A \textbf{user interface} designed for the infrastructure managers to perform propagation operations;
\end{itemize}
Perhaps most importantly, we hope that our approach can be utilized and adapted in future VIS efforts in emergency situations.

\section{Related Work}
\subsection{Automatic Visualization}
Zhu et al.~\cite{Zhu:2020:VI} presented a review on automatic tools and systems for generating visualizations. In their review, they divided the tools and systems into four different types: (i) tools that require programming and visualization knowledge, such as D3.js~\cite{Bostock:2011:TVCG} and Vega-Lite~\cite{Satyanarayan:2017:TVCG}; (ii) tools that utilize a visual building step, e.g., Charticulator~\cite{Ren:2019:TVCG} and Lyra~\cite{Satyanarayan:2014:CGF}; (iii) systems and tools that are semi-automated are require some form of user interaction for generating visualizations, such as Voyager~\cite{Wongsuphasawat:2016:TVCG} and Show Me~\cite{Mackinlay:2007:TVCG}; and (iv) automatic visualization generation tools and systems that are designed for users who are not experts in programming or visualization, e.g., Text-to-Viz~\cite{Cui:2020:TVCG} and Click2Annotate~\cite{Chen:2010:VAST}. We extend this classification, contributing a novel infrastructure and visual design pipeline that takes a visualization created by a designer and semi-automatically propagates it across a large data infrastructure.

Brodlie et al.~\cite{Brodlie:2005:CGF} presented a survey on a range of visualization applications requiring infrastructural support. Building on the emergence of autonomic computing as a new research agenda at that time~\cite{Kephart:2003:C}, they envisaged the need for introducing adaptive and autonomic techniques for managing VIS infrastructures. They started a discussion around infrastructure requirements for such systems, which we contribute to here through our infrastructure design and algorithmic support for visualization. Grammel et al.~\cite{Grammel:2013:EuroVisSP} presented a short survey of visualization construction user interfaces systems dividing the systems into six different approaches of: visual builder, visualization spreadsheet, textual programming, visual dataflow programming, template editor, and shelf configuration. Our propagation pipeline approach is a novel extension of the ``visual dataflow programming'' approach in their classification.

Several tools for generating visualizations automatically have been explored in the literature, of which we give a brief overview. Mackinlay~\cite{Mackinlay:1986:TOG} presented one of the first tools that automatically generates visualizations of relational information, such as bar charts, scatter plots, and connected graphs. Their approach is ideal for domains with easily defined semantics, although is impractical for a problem domain as complex and ever-changing as the response to a pandemic. Mackinlay et al.~\cite{Mackinlay:2007:TVCG} described Show Me, user interface commands integrated into Tableau that provide automatic views during the visual design workflow, part of a user-centric approach to visualization generation.

Falconer et al.~\cite{Falconer:2009:CISIS} presented an approach for generating customized visualizations through ontology mapping. Sun et al.~\cite{Sun:2010:SG} demonstrated Articulate, a novel conversational approach to visualization generation. Their system combined natural language processing and machine learning methods to enable the translation of imprecise sentences provided by the user into explicit expressions, which then automatically create a visualization through a heuristic graph generation algorithm. This aimed to simplify the visualization process by allowing users to describe what they wanted to see, without needing to know how to implement the visualization themselves. Cui et al.~\cite{Cui:2020:TVCG} also explored a natural language approach to visualization generation; they demonstrated an automatic approach for generating infographics from natural language statements -- the statements are converted from simple proportion-related statistics to infographics using pre-designed styles.

Automatic visualization approaches can be extended beyond single visualizations to more narrative forms of information dissemination. Shi et al.~\cite{Shi:2021:TVCG} presented Calliope, a system for automatically generating visual data stories from a spreadsheet. Their system progressively generates story points using a Monte-Carlo tree search algorithm, then assembles these into a single data story. Tang et al.~\cite{PlotThread2021} described PlotThread, an AI-assisted system for designing storyline visualizations. Their system provides an AI agent that works alongside the user to collaboratively produce a visual design, one of many examples of AI-assisted visualization generation systems~\cite{wu2021survey}.

Users are generally trusting such (semi-)automatically produced visualizations~\cite{zehrung2021vis}. Using such systems can lower the entry costs to visualization by simplifying the design process for the end user, or by taking them out of the loop entirely. However, as we will discuss in Section~\ref{sec:Overview}, a new approach was needed when creating a VIS infrastructure where bespoke visualizations (e.g., for epidemiologists and modeling scientists) needed to be rapidly deployed across a significant data infrastructure in a fast and cost-effective way.

\subsection{Ontology-Supported Visualization}
Ontologies can be a powerful tool in a data infrastructure, supporting the creation of visualizations through their structured representation of data, concepts, and relations~\cite{chen2016pathways}. Ontologies and their encoded  knowledge may also need to be visualized and there are many techniques for doing this. In this paper we focus on the use of an ontology to support visual design, rather than visualizing an ontology. For an overview of the latter, see surveys by Katifori et al.~\cite{katifori2007survey} and Dud{\'a}{\v{s}} et al.~\cite{dudavs2018ontology}.

Carpendale et al.~\cite{carpendale2014ontologies} presented a viewpoint on ontologies in biological data visualization, taking inspiration from their widespread use in biology research. They reflected on the technical challenges of ontology-based visualization in this domain and identified promising future research directions for the visualization community. Among these were ontology-supported visualization generation, leveraging the structure of an ontology to simplify visual design and exploration.

Ontology-supported automatic visualization was explored by Gilson et al.~\cite{Gilson:2008:Eurographics}, who presented an pipeline approach that combined ontology mapping and probabilistic reasoning to produce new visualizations. Their SemViz system exemplified this process, using three ontologies to automatically visualize music chart data. Khan et al.~\cite{Khan:2016:AEI} used an ontology in an enterprise search system to capture search provenance, using the ontology to visualize collaborative search graphs.

Yu and Silva~\cite{yu2016visflow} presented VisFlow, a visualization framework where a data flow diagram is used to support exploration and visualization of tabular data. Whilst this work did not use an ontology, its use of structured data representation for visualization is relevant and shows the value in using such representations to support the creation of new visualizations. Their FlowSense system~\cite{yu2019flowsense} extends this with a natural language interface for editing the data flow diagrams. Its semantic parser with special utterance tagging and placeholders are used to allow generalization to different datasets and data flow diagrams, simplifying visualization creation for the end user.

Our work uses an ontology to support the creation of new visualizations. We leverage the ontology in our data infrastructure to support propagating visual designs across many datasets, through a search-and-review workflow. Our streamlined workflow reduces the time-cost and volunteer effort necessary to scale the RAMPVIS system and support its domain experts with new visual designs.






\begin{figure*}[t!]
    \centering
    \includegraphics[width=0.87\textwidth]{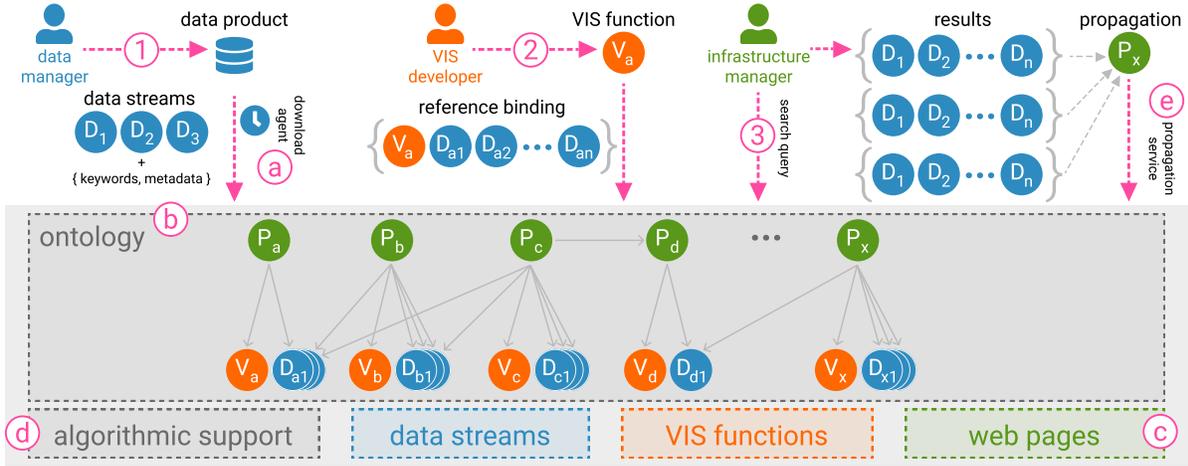}
    \caption{An illustration of our streamlined workflow, data infrastructure \euan{and VIS volunteer roles}. (1)~When new \textbf{data products} become available, a developer writes a manifest to extract \textbf{data streams}, add them to the \textbf{ontology (b)}, and keep data up to date via a \textbf{download agent (a)} \euan{that periodically queries the data product for new data}. (2)~When a new visualization is needed, a developer creates a \textbf{visualization function} (e.g., in D3.js) \euan{using a template} and binds it to reference data stream(s) in the ontology; \euan{this new visual design is accessible as a web page, visualizing the reference data}. (3)~To propagate \euan{the reference} visualization to other \euan{related} data streams, the infrastructure manager \euan{uses a search UI to find suitable data streams} then performs quality assurance on search results, produced by \textbf{algorithmic support (d)}. When a decision is made to propagate a visualization function to data stream(s), a \textbf{propagation service (e)} operates using the ontology. Propagated visualizations are immediately published as interactive \textbf{web pages (c)} for domain experts. \euan{The supplementary video walks through this workflow.}}
    \vspace{-4mm}
    \label{fig:Workflow}
\end{figure*}

\section{Problem Statement and System Overview}
\label{sec:Overview}
When the VIS volunteers first joined the SCRC effort for combating COVID-19 in May 2020, the SCRC data infrastructure was under development. From some example datasets in spreadsheets, we were overwhelmed by the amount of data. There were time series for different regions, genders, age groups, key indicators (e.g., number of tests, number of ICU patients, etc.), fatality locations (i.e., care homes, hospitals, etc.). There were different models being developed and tested, each of which would produce several time series for different transition states, and hundreds of multiples of such time series for uncertain or sensitivity analysis. Meanwhile, analytical algorithms, e.g., for comparing different datasets, were expected to result in even more datasets, potentially in a combinatorial manner. The scale of such an operation was already significant and would only grow over time.

Although the RAMPVIS generic support team consists of all VIS volunteers who offered to help engineer VIS systems and support others in the SCRC, there was only a very limited amount of programming resource: four VIS volunteers, about 2--15 hours per week per person, totaling 20--30 person-hours per week. Due to the nature of volunteering, we also had to assume that some VIS volunteers might become unavailable from time to time. This required us to devise a highly cost-effective approach, technically as well as operationally. We considered several optional approaches:
\begin{itemize}[noitemsep,nolistsep,leftmargin=*]
    \item \emph{Using an existing platform that would allow us to create plots and dashboards without programming.} \textcolor{gray}{\faThumbsDown}~We could not use this approach because (i)~the generic support team had to implement novel and nuanced visual designs produced by other teams~\cite{rampvis:2020:arXiv}; (ii)~we did not have any funds to purchase a server license and consultancy for database connection; and (iii) creating and managing numerous plots and dashboards would be challenging.
    \item \emph{Programming plots and dashboards with a UI for browsing suitable data.} \textcolor{gray}{\faThumbsDown}~We did not use this approach because (i)~it burdens domain experts with browsing hundreds or thousands of data streams; (ii)~it burdens each VIS volunteer with knowing all relevant data streams, programming interaction, and being responsible for the full data and visualization pipeline; and (iii)~it would demand substantial and consistent availability of resources.
    \item \emph{Programming with an advanced development framework (freeware).} \textcolor{gray}{\faThumbsDown}~We tried this approach for two weeks but stopped because (i)~we realized that only one person was knowledgeable about the suggested framework and libraries, and the learning curve for other volunteers was too high; and (ii)~we had doubts about how this would scale to numerous plots and dashboards.
    \item \emph{Programming reference plots and dashboards using a familiar platform and developing an infrastructure to propagate reference visual designs to work with all similar datasets.} \textcolor{darkgray}{\faThumbsUp}~After some discussion, we took this approach because (i)~each developer needed to cover a narrow spectrum of software development, facilitating a streamlined workflow; (ii)~because all developers knew D3.js, development could start immediately without the burden of retraining or raising funds; and (iii)~it reduces the development time for producing hundreds of plots and dashboards, ideal for a time-critical volunteer effort.%
\end{itemize}

Within two months the team developed several plots and dashboards and were able to propagate all plots with a single data stream. While this rapid development helped convince domain experts in the SCRC to make VIS the fourth pillar in combating COVID-19 (in addition to data, models, and policies), it also encouraged our team to develop more advanced propagation methods for more complex plots and dashboards.

\autoref{fig:Workflow} gives an overview of the main VIS infrastructure components and illustrates the overall workflow, from obtaining data, to creating plots and dashboards, to propagating these across all datasets and making them available to the domain experts. There are three main operations in the workflow overseen by VIS volunteers with distinct roles: \textit{data manager}, \textit{visualization developer}, and \textit{infrastructure manager}:
\begin{enumerate}[noitemsep,nolistsep,leftmargin=*]
    \item \textbf{Obtaining data streams---\autoref{fig:Workflow}~(1):} When a new \textbf{data product} needs to visualized, a \textit{data manager} writes a manifest to obtain \textbf{data streams}, assigns appropriate keywords, and enters metadata into the ontology \euan{via a simple web form}.
     \item \textbf{Writing VIS functions---\autoref{fig:Workflow}~(2):} A \textbf{VIS function} is an implementation that visualizes data streams in a single web page, \euan{e.g., as a plot or dashboard}. When domain experts require new designs, a \textit{visualization developer} \euan{is given a code template}, implements \euan{the visual design} and binds with reference data streams, \euan{creating a reference visualization accessible as a new web page}.
    \item \textbf{Propagating to other streams---\autoref{fig:Workflow}~(3):} An \textit{infrastructure manager} \euan{uses our search UI to find} data similar to the reference data in a VIS function then activates propagation for appropriate results, \euan{propagating that design across numerous data streams}.
\end{enumerate}

These operations that are performed by VIS volunteers are supported by key technical components of the infrastructure:
\begin{itemize}[noitemsep,nolistsep]
    \item[a.] \textbf{Download agent---\autoref{fig:Workflow}~(a):}
    Periodically obtains data streams from the SCRC data infrastructure \euan{without human input}. 
    \item[b.] \textbf{Ontology---\autoref{fig:Workflow}~(b):} A knowledge representation of data streams, VIS functions, and their combinations as web pages. Supports search activities during propagation and general use for supporting the domain experts' activities.
    \item[c.] \textbf{Web-based visualization service---\autoref{fig:Workflow}~(c):} Hosts numerous visualization plots and dashboards in a scalable manner.
    \item[d.] \textbf{Algorithmic support---\autoref{fig:Workflow}~(d):} Supports propagation by searching for similar data streams registered in the ontology, ranking similarity levels, and ordering the search results.
    \item[e.] \textbf{Propagation service---\autoref{fig:Workflow}~(e):} Propagation is supported by a specialized UI for searching and activating propagation.
\end{itemize}

Components (a)--(c) are essential parts of the VIS infrastructure, but were enhanced to support components (d) and (e) which were specially developed for propagating visualizations. \aar{All components have been developed as open-source software and the code is available \cite{RAMPVIS-Onto-UI:2021:git, RAMPVIS-API:2021:git, RAMPVIS-UI:2021:git}. The design and development of (d) and (e) took 200$\sim$400 person-hours.}

\subsection{VIS Infrastructure Ontology}
\label{sec:overview:ontology}
\begin{figure}[ht!]
    \centering
    \includegraphics[width=0.65\columnwidth]{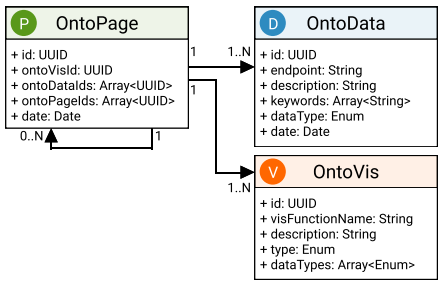} 
    \caption{Schema showing the core classes in our \euan{ontology} and their relationships to each other.} 
    \label{fig:ontology-schema}
    \vspace{-2em}
\end{figure}

%
In our VIS infrastructure, an ontology is used to organize data and visualizations, and to support propagation.\ct{We argue that an ontology is a suitable method to deal with the sheer number of diverse data streams and visualisations that we have in our problem context. Considering also that the set of data and designs are constantly evolving over time, a well-designed ontology not only forms the basis of the infrastructure design but also provides versatile and robust means to manage the propagation process.}  \autoref{fig:ontology-schema} shows a schema of our ontology. \textit{OntoData}, \textit{OntoVis}, and \textit{OntoPage} are the three main classes, representing data streams, VIS functions and web pages, respectively (as in \autoref{fig:Workflow}). The ontology is implemented as a graph data structure: objects are mapped to nodes, relationships between objects are mapped to edges, and directed edges distinguish start/end nodes. In the following sections, we discuss data streams (Section~\ref{sec:DataStreams}), visualization functions (Section~\ref{sec:VisFunctions}), and our novel propagation process for binding these to create new visualizations (Sections~\ref{sec:UI} and~\ref{sec:algorithm-support-for-vis-function-propagation}).

\section{Data Stream}
\label{sec:DataStreams}

%
Data streams are units of data in our infrastructure, with associated keywords and metadata. Our infrastructure provides access to data streams via a RESTful API. The OntoData class in our ontology stores registered data streams and their attributes (see \autoref{fig:ontology-schema}). From these attributes, \textit{endpoint}, \textit{description}, and \textit{keywords} are most relevant: the endpoint is an RESTful API endpoint used for accessing the data; the description should describe the data and is used for search and discovery; and the keywords describe the stream contents, and are used for searching, grouping, and propagation.

We create data streams using data from the SCRC infrastructure~\cite{scrcdata}, which includes a wide range of COVID-19 data. SCRC data is organized into data products (e.g., testing, hospital, mortality), each further divided into components (e.g., deaths per council area, deaths per age group). To fetch data, a VIS volunteer writes a download agent to extract data from the SCRC infrastructure and transform it if needed (e.g., to normalize per 100,000 people). Relevant keywords are assigned, then the stream is registered in our ontology. Data products are updated periodically by SCRC, so download agents update data daily. 
 
%
Data streams can be searched using the OntoData description and keyword attributes, in $O(n)$ time. To improve efficiency, the Lucene~\cite{ese:lucene:web} open-source text search engine is used to create an inverted index~\cite{Manning2008:Book} that maps description and keywords attributes to their matching instances. The inverted index is a hash map-based data structure, allowing searches in $O(1)$ time complexity. An indexing agent periodically scans the ontology database logs for changes and keeps the index updated.

Description fields are broken into individual words, $2$-grams and $3$-grams for indexing. The words and their $n$-grams support partial matching in a search and allow hints while typing queries, simplifying infrastructure manager operations. For example, a description with ``positive cases'' would have all its components indexed: ``positive'' and ``cases''; a query for either word would return this string. Keywords are indexed as-is and are not broken down for partial matching.

\section{Visualization Functions}
\label{sec:VisFunctions}

\subsection{Visual Design Workflow}
\label{sec:visual-design-workflow}
In our infrastructure, a VIS function is an implementation of a visual design (e.g., plots, dashboards). These functions are created by a VIS volunteer using familiar libraries (in this instance, D3.js~\cite{D3}). The OntoVis class in our ontology stores visualization functions and their attributes (see \autoref{fig:ontology-schema}). From these attributes, \textit{visFunctionName} is most important: this is an identifier for a Javascript function that will create an interactive visualization for given data streams.

A visualization function will be linked to a set of data streams and rendered on a web page for domain experts and visualization viewers. The OntoPage class in our ontology represents a web page, establishing a link between one VIS function (an OntoVis instance) and a set of data streams (one to many OntoData instances), as shown in \autoref{fig:ontology-schema}. Note that each OntoPage instance may also be linked to other OntoPage instances; for example, a dashboard may show several plots and be linked to their individual OntoPage instances.

When a VIS developer needs to implement a new visual design (e.g., to support domain expert requirements), they liaise with the infrastructure manager, who will: (1)~create an OntoVis instance, registering the function in the ontology; then (2)~create an OntoPage instance, by binding the new OntoVis instance to an appropriate set of reference data streams (i.e., instances of OntoData). The new OntoPage instance results in a web page `template' \euan{with placeholder code} that the VIS developer can use to implement the visual design. The reference data serves two purposes: providing test data for the developer \euan{to support implementation}, and providing an initial binding between the VIS function and data streams in the ontology.

Our visual design workflow is based on the core concept that visualization implementation is decoupled from the rest of the infrastructure. When the infrastructure manager creates the reference `template' in the ontology, it appears in the development instance; VIS developers then implement their VIS function and push their code to the repository, making it available immediately. This streamlines the process because developers do not need to know about, or work directly with, the underlying data infrastructure. This makes our approach suited for volunteering operations. It also facilitates efficient propagation for producing numerous plots and dashboards with minimized time-cost.

\subsection{Creating Web Pages}
\label{sec:creating-web-pages}

Each OntoPage instance in our ontology yields an interactive web page that domain experts can use to access a plot or dashboard. Our implementation uses the Flask Jinja template engine~\cite{jinja2} to extract information from an OntoPage instance and generate a web page:

\begin{enumerate}[noitemsep,nolistsep,leftmargin=*]
    \item Title and description are extracted from OntoPage attributes;
    \item HTTP requests fetch data from each OntoData API endpoint;
    \item The visualization function is identified from the OntoVis instance and its JavaScript object is retrieved from an object factory;
    \item The visualization function is called with the fetched data streams, rendering the visualization on the page;
    \item If the OntoPage is linked to other OntoPages, these attributes create hyperlinks to their web pages.
\end{enumerate}

\begin{figure}[th!]
    \centering
    \includegraphics[width=0.91\columnwidth]{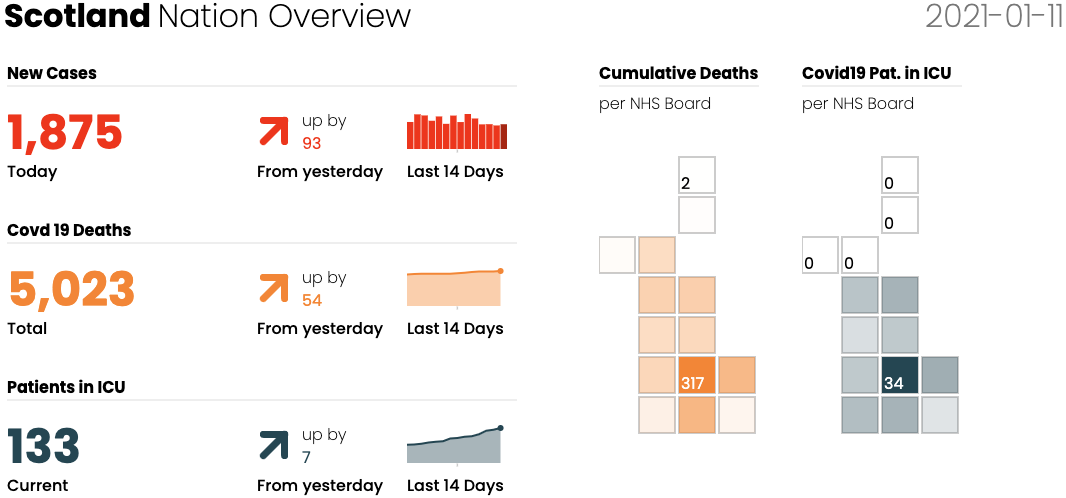}\vspace{1em}
    \includegraphics[width=0.91\columnwidth]{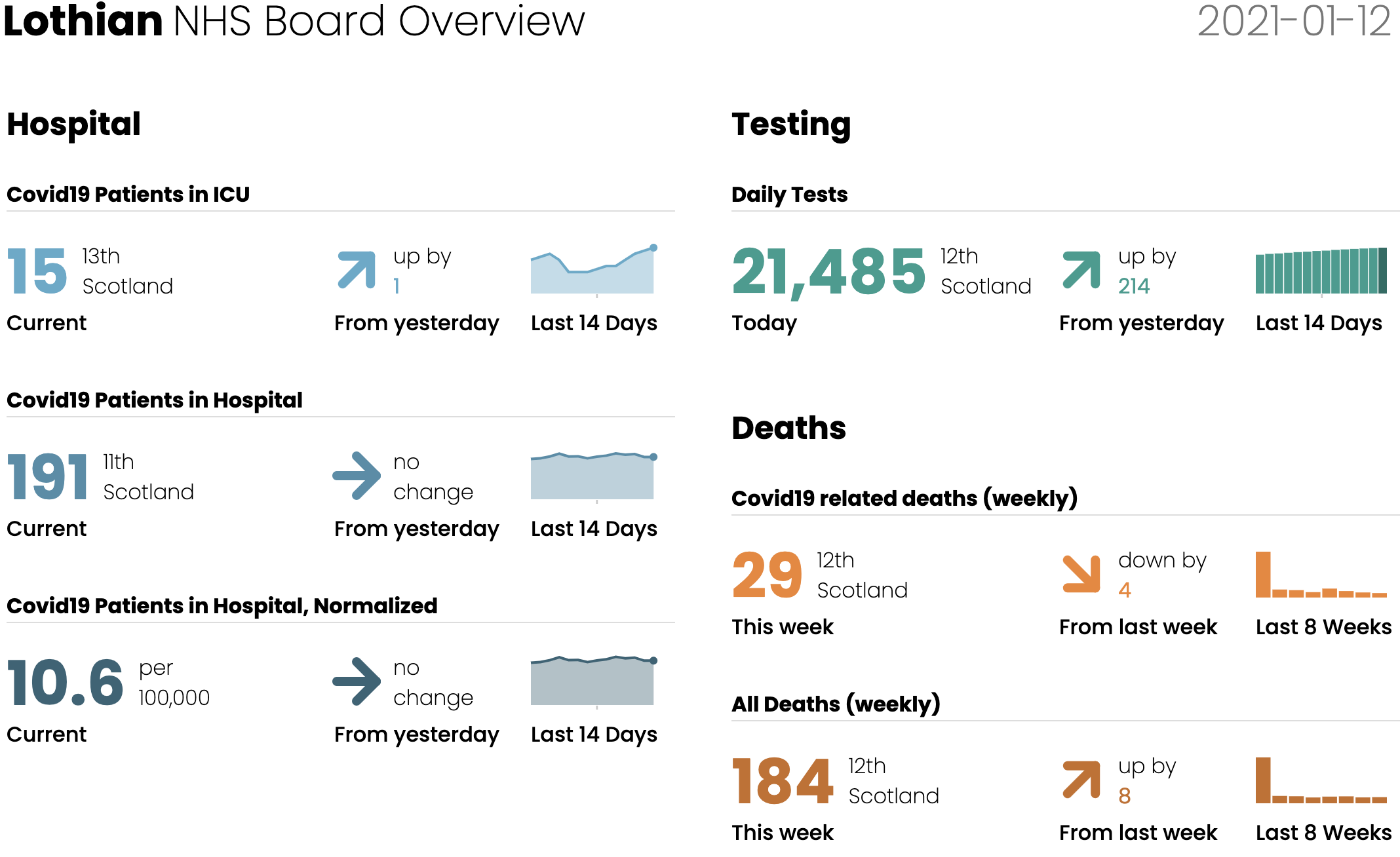}
    \vspace{-4mm}
    \caption{Example dashboards for Scotland and a regional health board. The nation cartogram links to dashboards for the health board regions.}
    \vspace{-1em}
    \label{fig:dashboards}
\end{figure}

\subsection{Template-Based VIS Function Design}
Our template-based approach to visualization implementation supports a variety of visual designs, which we categorize as \textit{plots} (single visualizations) or \textit{dashboards} (bespoke composite visual designs with multiple plots, annotations, etc). We briefly discuss how these are implemented in our infrastructure.

\textbf{Plots}---We implement a variety of visualizations (e.g., line chart, bar chart, area chart, chord diagram, matrix, map) for many different data types (e.g., time series, cumulative time series, matrix, geographic data). Our infrastructure and propagation process is agnostic to the detailed visualization design and implementation, so is able to accommodate all of the domain experts' visualization needs.

Plots can visualize multiple data streams, as illustrated in \autoref{fig:teaser}. In that example, a stacked bar chart shows weekly location of death in a region of England, with a unique data stream for each of the six locations. The relationships between the VIS function and its six data streams are in the ontology (via the OntoPage instance). By implementing plots in this way, a single VIS function can be propagated across hundreds of data streams. As shown in \autoref{fig:teaser}, we created this plot with reference data streams from Oxford (left plot), then propagated that single function to all regions in England (e.g., Birmingham, City of Bristol, and Westminster in the right plots). When this propagation occurs, each data stream needs to be replaced by the appropriate data stream for the other regions (e.g., replacing Oxford deaths in hospital with Birmingham deaths in hospital).

Plots may also have links to other plots (OntoPage$\rightarrow$OntoPage). For example, a plot showing small multiples of COVID-19 patients in ICU of all national health boards can link to each health board. 

\textbf{Dashboards}---As well as individual plots, our approach is also capable of supporting composite dashboards with several complementary plots drawing from different data streams. Dashboards summarize important data about multiple data streams, such as current data and trends from recent days. Dashboards serve the following purposes: (i) they provide quick access to frequently-used plots; (ii)~they provide rapid access to critical information to inform daily decision making (e.g., deciding to call an emergency meeting, or checking if model predictions match current data); and (iii) they avoid unnecessary search activities, simplifying decision making and review processes.

\autoref{fig:dashboards} shows two of our implemented dashboard visualizations. These summarize data from all of Scotland (top) and one region of Scotland (bottom). Each dashboard has been carefully designed to give an immediate and accurate overview over relevant data, satisfying domain expert requirements. Importantly, each component in a dashboard is linked to the corresponding web page for the individual plot: a viewer can click any of the numbers, arrows, or trend charts to open the full detail view. The cartogram in the nation overview (top) shows each of the NHS Scotland health boards; each region in the cartogram is linked to the dashboard for that health board region, so that clicking a region will lead to regional dashboard, e.g., NHS Lothian in \autoref{fig:dashboards} (bottom).

In total we designed five dashboards each centering about a specific topic such as: a particular region a nation in the UK, hospitals, schools, places of death. Using our propagation mechanism we can propagate these dashboard designs to all Scottish regions, ensuring that each data stream is replaced by the appropriate data stream for the other region.

Dashboards are implemented using the same process as individual plots. Each dashboard has an OntoPage instance, a single VIS function that produces the visual design and page layout, and a set of all associated data streams. Each component in the dashboard is linked to its individual visualization web page: e.g., New Cases in the Nation Overview is linked to the web page visualizing daily cases and the cartogram regions in Nation Overview are linked to the web page for the regional dashboards. These links are stored in the OntoPage attributes and are linked by the VIS function. Consequently, propagating dashboards is more complex than propagating simple visualizations, as the data streams and links need to be correctly matched.

\section{Propagation Service}
\label{sec:UI}

In our ontology (\autoref{fig:ontology-schema}), OntoPage objects create a binding between visualization functions (OntoVis), data streams (OntoData) and other web page links (OntoPage). VIS functions can be propagated to other relevant data streams and links, a process that results in new OntoPage instances with the same VIS function and a new set of data streams and links. This is a novel aspect of our ontology-based approach to visualization, as existing VIS functions can be used to visualize numerous data streams, without any action from VIS developers.

Propagating a VIS function to generate plots and dashboards for other data streams is not straightforward. This requires actions from the infrastructure manager to ensure all appropriate data streams are correctly mapped and linked in the propagated visualizations. Whilst visual designs in our system are implemented by volunteer VIS developers, the infrastructure manager is a volunteer responsible for overall infrastructure management and visualization propagation. The infrastructure manager faces several challenges: (i)~there are numerous data streams in the infrastructure and knowing which streams are available is difficult; (ii)~some plots and dashboards have multiple data streams and links that need to be correctly matched, but searching for matching data streams and links is a group-based multi-criteria decision; (iii)~when there are many possible matching results, quality-assurance is a mission-critical and demanding task.

Our propagation workflow (\autoref{fig:Workflow}) has two tasks, carried out by the infrastructure manager: first, they need to formulate a query for data streams that can be part of a sensible binding with the chosen visualization; second, they must perform `quality assurance' by reviewing search results, to determine whether to propagate the visualization and have it published. \euan{Our system has a search user interface to support these two tasks, which} we now discuss separately.

\subsection{User Interface for Search and Ranking}
\label{sec:ui:search}

When a VIS function needs to be propagated, the infrastructure manager first needs to search for appropriate data streams. We have a significant number of data streams (e.g., dozens of metrics stratified by dozens of local authority regions) and reviewing every permutation of data streams for a new visualization will be time-consuming and impractical. The search process and user interface aims to help the infrastructure manager find good candidates for propagation. An effective search interface will help with quality assurance by reducing the potential for inappropriate bindings, reducing the volunteer time-cost. It will also help reduce the workload required to disseminate new visualizations, especially as the system scales with new data streams and more complex visualizations (e.g., dashboards with several plots).

\begin{figure}[t!]
    \centering
    \includegraphics[width=0.95\columnwidth]{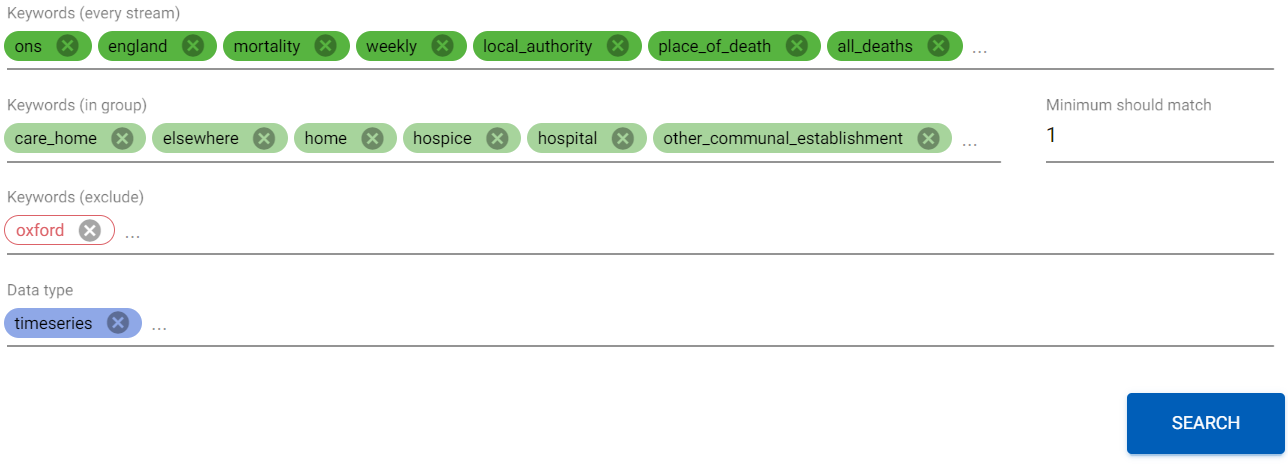}
    \caption{Screenshot of the search user interface with an example query \euan{consisting of keywords} for six data streams about weekly mortality rates in England. \euan{Keywords can be quickly added to a query via the search UI.}}
    \label{fig:search_ui}
    \vspace{-4mm}
\end{figure}

Search and result ranking operates on keywords in the ontology. Since every VIS function is defined with reference data streams, our system extracts keywords from those references and uses them to search. When starting a new search for a chosen visualization function, the infrastructure manager is shown the keywords for the \euan{reference} visualization and data streams (as in \autoref{fig:search_ui}). There are four search bars for building a query: for identifying keywords that (1)~must appear in every data stream, (2)~must appear at least once within a group of results, (3)~must not appear, and (4)~for limiting data types.

Clicking on a keyword in the \euan{list of} reference data streams will add it to the first search bar (i.e., must appear in every stream) and subsequent clicks will move them to the next bar, cycling through the three keyword criteria. Keywords are shown with a colored background, both in their original location and in the search bar (e.g., \autoref{fig:search_ui}): dark green means the keyword is in every stream, pale green means a keyword must appear at least once in a group of results, and a red border and text means a keyword must be excluded. Keywords may be excluded because they are expected to vary or be omitted (e.g., when propagating a dashboard for one region to other regions). Data type filters are shown in their own search bar with a blue background.

Overall, this user interface supports query construction by using a visualization's reference data binding as the `template' for building queries from ontology keywords. This reduces the need for text entry and ensures keywords are entered correctly. The user interface visualizes the search parameters \textit{in situ} in the reference visualization by highlighting keywords in its data stream(s), \euan{helping the infrastructure manager} verify the search criteria \euan{are formulated correctly}. Search results are then presented in ranked order. Parameters for the ranking algorithm can also be adjusted via the UI if necessary, e.g., to specify the required number of matching keywords within a group of streams.

\subsection{User Interface for Quality Assurance}
\label{sec:ui:qa}

Search results are presented \euan{with grouped and highlighted keywords} so the infrastructure manager can \euan{see at a glance} if a visualization function should be propagated \euan{to a set of results}. This quality assurance is necessary to ensure that visualization functions are only propagated and published if they are appropriate for the underlying data. Importantly, this only happens once for each visualization function and data stream permutation: once a visualization is propagated and a new binding has been established, no further review is required. This allows the system to scale, without the need for frequent and time-consuming quality assurance. Downloader agents and propagation agents ensure visualizations on web pages are automatically kept up to date.

Having a human in the loop is vital as this is a complex decision process. The infrastructure manager needs to evaluate each search result. For a visualization for a single data stream, this is a straightforward check to decide if the data types match and the data would make sense for that visualization. For complex visualizations of multiple data streams (e.g., regional or national dashboards, multi-series plots), this requires more nuance. Data streams, types, and ontology keywords must be compared with the vis.\ function reference data streams, to decide if propagation makes sense for each permutation of streams. Propagation only takes place once; i.e., if a visualization is propagated to a set of data streams, then that result is not shown in future.

Our search result user interface helps the infrastructure manager make these decisions \euan{using keyword grouping and highlighting}. Whilst our search and ranking process reduces inappropriate matches, there is still a considerable amount of results to review. A good user interface thus has a large positive impact on reducing workload, which is important when dealing with data and visualizations of this nature.

\begin{figure}[t!]
    \centering
    \includegraphics[width=0.48\textwidth]{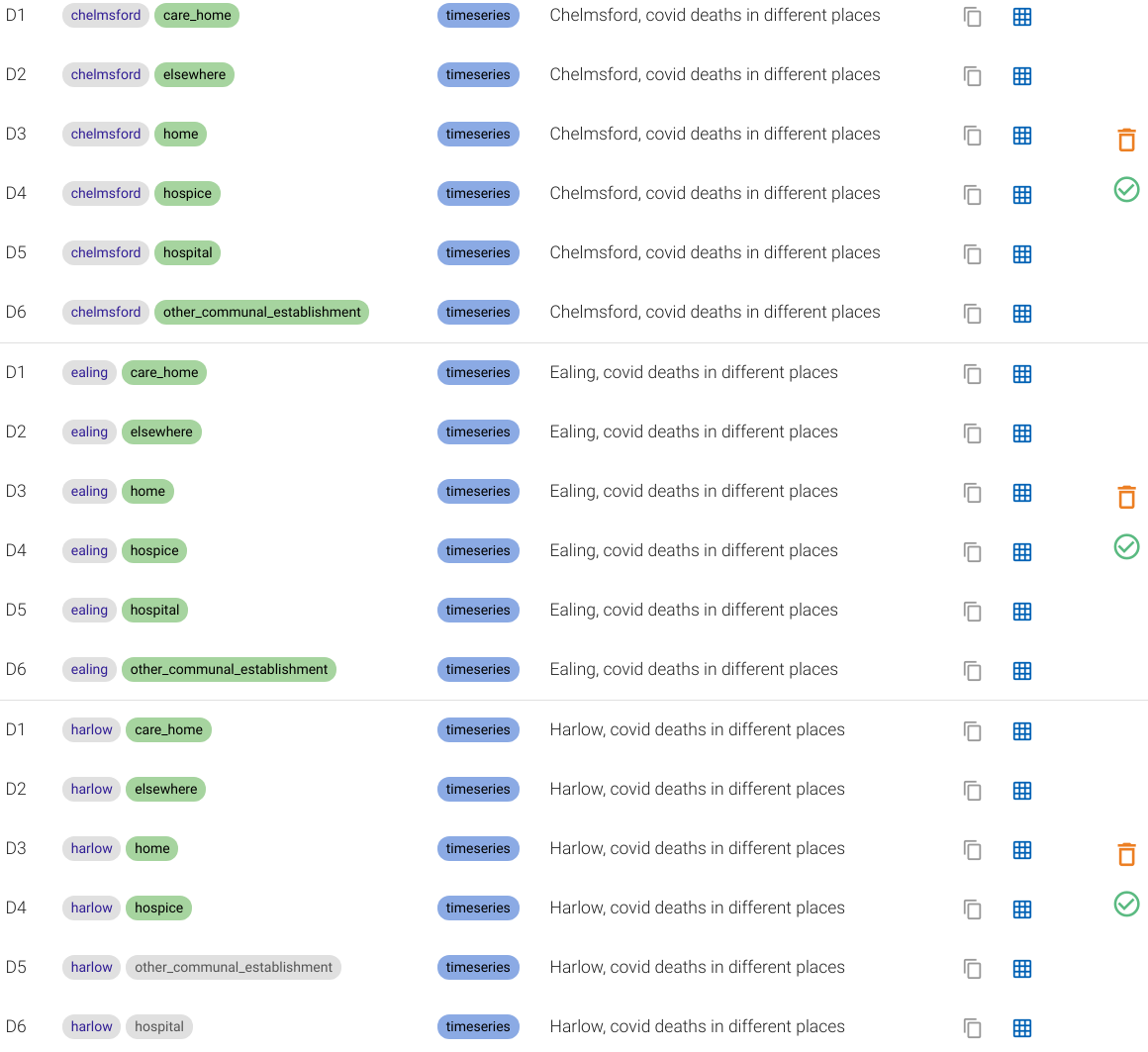}
    \caption{Screenshot of the search results user interface for propagation quality assurance, showing three results with six data streams (for the \autoref{fig:search_ui} query). Keywords are grouped and highlighted by match type, so \euan{the infrastructure manager can glance at the results keywords and decide quickly if propagation should be activated (via the tick icon).}}
    \label{figure:qa_ui}
    \vspace{-4mm}
\end{figure}

\autoref{figure:qa_ui} shows how search results are presented. Each result is a set of data streams that satisfy the query constraints. Each data stream is shown with its keywords, data type, and description. Keywords have the most significant influence on infrastructure manager's decisions about whether to propagate a visualization to a set of data streams, so we structured the keyword presentation to facilitate efficient comparison between the visualization reference data keywords and the result data stream keywords. If keywords appear in all reference and result streams they are not shown, since \euan{their presence} is implicit from the query. Keywords that do not appear in the reference streams are shown first with a gray background to help identify differences. If keywords match a query term and appear in the correct order, they are shown with a pale green background. By grouping and presenting keywords like this, the infrastructure manager can decide if propagation should occur.

As an illustrated example, \autoref{figure:qa_ui} shows three search results for a query (shown in \autoref{fig:search_ui}) for a stacked bar chart with six data streams (shown in \autoref{fig:teaser}). Each result is a set of six data streams that meet the query constraints. The first column shows keywords that do not match the query terms, \euan{highlighted in gray}. In the three search results, these unmatched keywords are for three other regions of England, \euan{so it is expected that they do not match query terms, as we want to propagate this design to other regions}. In this example, the visualization shows data per region, so the infrastructure manager will see that these data streams are \euan{grouped by one region} and valid for this visual design.

The second column shows keywords that matched query terms. In the first two results, all keywords in the second column are green, showing a complete match: these keywords appear in the query and correspond with the ordering in the visualization reference. Since these two results are correctly grouped by region and the data streams match, the infrastructure manager would choose to propagate (using the green check-mark icon). In the third result, all six keywords match the query terms, but the last two appear in the incorrect place (and are highlighted gray to show this). Propagation should not take place for this result, as some data would appear incorrectly. 

\section{Propagation Technical Infrastructure}
\label{sec:algorithm-support-for-vis-function-propagation}

\begin{figure}[t!] 
    \centering
    \includegraphics[width=0.65\columnwidth]{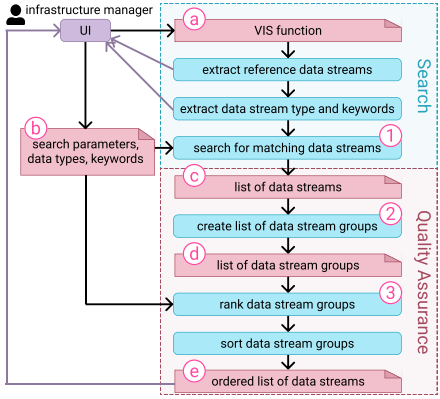}
    \caption{A flowchart showing the search and ranking process. When the infrastructure manager selects a \textbf{VIS function~(a)}, its reference data streams and their attributes are retrieved and shown in the search UI. They then construct a \textbf{query~(b)} \euan{by clicking on, or manually entering, keywords} then search. Search algorithms find matching \textbf{data streams~(c)}, create \textbf{stream groups~(d)}, sort them, then present an \textbf{ordered list (e)} through the results UI, \euan{highlighted to support fast visual scanning}.}
    \label{fig:search-data-strteam-algorithm-flowchart}
    \vspace{-4mm}
\end{figure}

Our propagation process is supported by a technical infrastructure of search, grouping, and ranking algorithms. \autoref{fig:search-data-strteam-algorithm-flowchart} shows the operations and algorithms underlying the search and quality assurance interfaces described in the previous section. When the infrastructure manager is ready to propagate a VIS function, its reference data stream keywords and metadata are extracted from the ontology.

\begin{figure*}[h!] 
    \centering
    \includegraphics[width=0.85\textwidth]{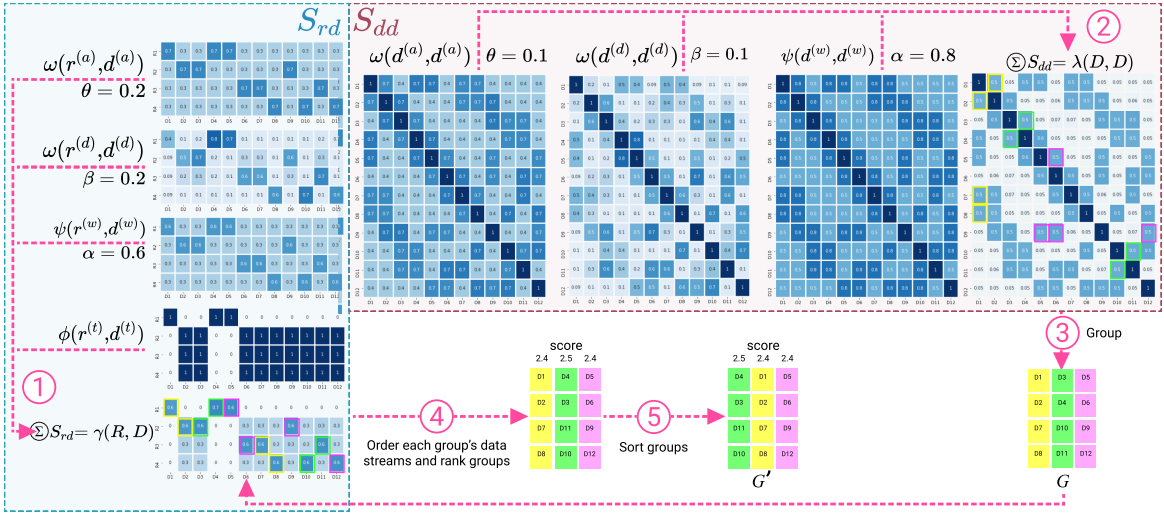}
    \caption{
    Illustrates an example grouping, ordering and ranking workflow with four reference data streams ($\mat{R}_1$, $\mat{R}_2$, $\mat{R}_3$, $\mat{R}_4$) and 12 discovered data streams ($\mat{D}_1$, $\mat{D}_2$ ... $\mat{D}_{12}$).
    (1)~We derive a similarity matrix $\mat{S}_{rd}$ measuring similarity between reference data streams and discovered data streams. 
    First we derive feature vectors from reference data streams (e.g., API endpoint $\vec{r}^{(a)}$, description $\vec{r}^{(d)}$, keywords $\vec{r}^{(w)}$, and data type $\vec{r}^{(t)}$) and discovered data streams (e.g., API endpoint $\vec{d}^{(a)}$, description $\vec{d}^{(d)}$, keywords $\vec{d}^{(w)}$, and data type $\vec{d}^{(t)}$). Next, we compute a pairwise similarity matrix: $\omega({\vec{r}^{(a)}, \vec{d}^{(a)}})$), $\omega({\vec{r}^{(d)}, \vec{d}^{(d)}})$, $\psi(\vec{r}^{(w)}, \vec{d}^{(w)})$, and $\phi(\vec{r}^{(t)}, \vec{d}^{(t)})$. Finally, we aggregate by taking a weighted average of the matrices.
    (2)~We derive the similarity matrix $\mat{S}_{dd}$ for discovered data streams using a similar process.
    (3)~Grouping algorithm group similar data streams into uniform groups.
    (4)~Data streams are ordered within each group, to match the reference stream order, then compute ranking scores.
    (5)~Sort groups by ranking score.
    }
    \label{fig:illustration-grouping-ordering-datastream}
    \vspace{-4mm}
\end{figure*}

Those reference data streams, keywords, and metadata are presented in the search UI (Section~\ref{sec:ui:search}), to help the infrastructure manager construct their query. Keywords are assigned one of three categories: must appear in every stream, must appear at least once within a group of streams, or must be excluded. Additional keywords and search terms can also be provided via the search UI.

When the query is ready, a series of algorithms process all data streams in the ontology. There are three algorithms: \textbf{Searching} (\autoref{fig:search-data-strteam-algorithm-flowchart}--(1)), \textbf{Grouping} (\autoref{fig:search-data-strteam-algorithm-flowchart}--(2)), and \textbf{Ranking} (\autoref{fig:search-data-strteam-algorithm-flowchart}--(3)); the final sorting process is trivial based on ranks. By searching then sorting results in the quality assurance UI (Section~\ref{sec:ui:qa}), the infrastructure manager is shown the best candidates for propagation at the top of the search results.

In the following sections we discuss the Searching, Grouping, and Ranking algorithms. We focus on plot propagation as a simple example. When propagating a plot that consists of several data streams, the priority is to find matching data stream groups such that the semantic ordering of streams is correct (e.g., so that the correct categories of data appear in the same order). Propagating a dashboard is more complex, because web page links (i.e., OntoPage instances) also need to be grouped and ordered correctly. This is a more complicated process, which we describe in the supplementary material.

\subsection{Data Stream Search} 
\label{sec:Searching}

Keywords are important when searching for data streams, as they are used to identify similar data streams appropriate for the chosen visual design (e.g., the plot or dashboard being propagated). Once a query is constructed in the search UI, it is converted into the declarative query language DSL~\cite{elasticquerydsl:2021:web}, for searching the database underlying the ontology.

As an illustrated example, consider \autoref{fig:teaser}. This shows stacked bar plots of regional weekly mortality data in England, which is split into six location types (care home, communal establishment, elsewhere, home, hospice, hospital) and provided for 336 regions. The Reference Visualization is for the Oxford region. Underlying this is an OntoPage instance, linked to the OntoVis instance for the VIS function and six OntoData instances for Oxford's mortality data streams.

Suppose the infrastructure manager wishes to propagate this plot to the other 335 regions of England. They construct the query shown in \autoref{fig:search_ui} which specifies: keywords that must appear in every data stream (e.g., england, weekly, mortality, place\_of\_death, etc), keywords that should appear in at least one data stream in the group (i.e., keywords for each place of death); keywords that should be excluded (i.e., for the Oxford region). It also specifies stream data type (i.e., time series).

Let the reference data streams for the visualization function be $R_1, R_2, R_3, \ldots, R_k$ (where $k=6$ reference streams). We search for all data streams in the ontology that match the search criteria using the specified keywords. This results in a set of $m$ discovered data streams $D_{(i,1)}, D_{(i,2)}, \ldots, D_{(i, m)}$. In total, there are $m=335$ sets of data streams (and each set containing $k$ data streams) discovered by our search algorithm, for all regions of England excluding Oxford.

\subsection{Data Stream Grouping} 
\label{sec:Grouping}

The reference data streams for a visualization function form a group, where the order of data streams is important. Let the reference data streams be $R_1, R_2, \ldots, R_k$; we would like to create similar groups that match this. Discovered data streams (total $n$) therefore need to be grouped in a similar way to the reference streams. In our example from \autoref{fig:teaser}, there are over 300 groups taken from thousands of data streams matching the query and, inevitably, there will be unwanted streams in the ontology that are discovered by the search algorithm.

Our grouping algorithm constructs groups from discovered data streams, aiming to maximize similarity with the reference stream group, outlined in \autoref{fig:illustration-grouping-ordering-datastream}. To do this, we compute two similarity matrices, $\mat{S}_{rd}$ and $\mat{S}_{dd}$, which are of size $k \times n$ and $n \times n$ respectively. We compute a similarity measure $\gamma(R_i, D_j)$ ($i=1..k, j=1..n)$ for each discovered data stream $D_j$. We compute another similarity measure $\lambda(D_u, D_v)$ ($u=1..n, v=1..n)$ for each pair of discovered streams $D_u$ and $D_v$.

The similarity functions $\gamma()$ and $\lambda()$ consider the similarity between data type, keywords, \sk{API endpoint} and the description field. \sk{API endpoint} and description similarity is computed using a text comparison algorithm, data type similarity is simple string matching function, and keyword similarity is based on comparison of sets. The three similarity measurement algorithms and computation of $\mat{S}_{rd}$ and $\mat{S}_{dd}$ are provided as Supplementary Material.

After computing the similarity matrices, the grouping algorithm examines the set of discovered data streams $D_1, D_2, \ldots, D_n$ and finds all permutations that meet a set of grouping requirements. Given a subset of data streams $[D'_1, D'_2, \ldots, D'_k] \subset [D_1, D_2, \ldots, D_n]$, the grouping requirements are defined using both similarity matrices $S_{rd}$ and $S_{dd}$:
\vspace{-1em}
\begin{align*}
     &\frac{1}{k}\sum_{i=1}^k \gamma(R_i, D'_i) > T_\text{group} \quad \land \quad \forall i=1..k,\, \gamma(R_i, D'_i) > T_\text{stream}\\
     \land \quad &\frac{2}{k(k-1)} \sum_{i=1}^{k-1}\sum_{j=i+1}^k \lambda(D'_i, D'_j) > S_\text{allpair}\\
     \land \quad &\forall i=1..k, j=1..k, i \neq j,\, \lambda(D'_i, D'_j) > S_\text{pair}
\end{align*}
\noindent where $T_\text{group}$, $T_\text{stream}$, $S_\text{allpair}$, and $S_\text{pair}$ are control parameters defined by the infrastructure manager. As a result, the grouping algorithm gives $m$ groups $G_1, G_2, \ldots, G_m$, each with $k$ data streams.

\textbf{Grouping Data Streams.} In Section~\ref{app:algorithms-for-grouping-and-ordering} of the supplementary material, we described two grouping algorithms, each with certain trade-offs. Algorithm \ref{algo:grouping-bruteforce} is based on a brute-force approach, which iterates through each row of the similarity matrix $\mat{S}_{dd}$ to find the $k$ closest elements, and keeps iterating until all $m$ groups are discovered. This works well in a situation where there exists exactly $k$ closest elements in each row.

Algorithm \autoref{algo:grouping-spectral} is based on a graph spectral method described in \cite{Leskovec2020:Book}. Graph spectral methods are applied to divide a graph's closest vertices into equal size components. The similarity matrix $\mat{S}_{dd}$ can be seen as an adjacency matrix of a weighted undirected graph: each element of $\mat{S}_{dd}$ represents a node of the graph, and the similarity between any two elements is the weight of an edge between them. This algorithm takes $\mat{S}_{dd}$ as an input and returns $m$ different groups $\mat{G}$, and each group containing $k$ data streams. This algorithm performs efficiently when the matrix is sparse and there exists only a small number of clusters.

The infrastructure manager can select which grouping algorithm to use when searching. Each performs better in different contexts and the decision to change will be ad hoc based on the results. For example, if there are a lot of data streams in the query then better results may be obtained by switching algorithm, whereas the graph spectral method will typically perform better for smaller numbers of groups.

\textbf{Ordering Data Streams.} In Section~\ref{app:algorithms-for-grouping-and-ordering} of the supplementary material, we outline the process where streams within a group are ordered (\autoref{fig:illustration-grouping-ordering-datastream}---(4)). We iterate over each column of the derived matrix, $\mat{G}$, to sort its streams, such that they match the reference group data streams based on the degree of similarity in similarity matrix $\mat{S}_{rd}$.

\subsection{Group Ranking} 
\label{sec:Ranking}

The ranking algorithm assigns a score to each group of data streams to indicate the likelihood that it may be suitable for propagation. The ranking score is computed based on both similarity matrices $S_{rd}$ and $S_{dd}$ described before. Given a group $G_a$ $(a=1..m)$ that contains $k$ data streams $[D^a_1, D^a_2, \ldots, D^a_k]$, the ranking score is defined as:
\vspace{-1em}
\[
    S(G_a) = \frac{W}{k}\biggl( \sum_{i=1}^k \gamma(R_i, D^a_i) \biggr) + \frac{2(1-W)}{k(k-1)}\biggl( \sum_{i=1}^{k-1}\sum_{j=i+1}^k \lambda(D^a_i, D^a_j) \biggr)
\]
\noindent where $W$ is a control parameter in the range of [0, 1] for controlling the contributions of the two types of similarity measures. Once a ranking score is assigned to each group, sorting them is trivial to determine the presentation order. The $m$ sorted groups, each with $k$ data streams, are sent to the results UI for quality assurance (Section~\ref{sec:ui:qa}).

%

\section{\aar{Evaluation}}


\aar{
We conducted a qualitative evaluation of the propagation interface and workflow during the development process, to reflect on our design and formatively evaluate the workflow effectiveness. There were six participants: two visualization researchers (our infrastructure and data managers respectively, co-authors), two experienced analytics developers (Power BI, Salesforce Einstein Analytics, and Tableau), one software developer, and one postgraduate student specializing in visualization. Each session began with a tutorial, after which participants were asked to complete two propagation tasks using real COVID-19 data and existing visualization functions. The session ended with a semi-structured interview. Each session lasted 60--90 minutes.
}

\aar{
We asked participants to reflect on the overall process of propagating plots to new data streams. As visualization researchers and developers, all understood the problem that propagation addresses. All participants recognised the time that propagation could save: some noted the time needed to search and group sets of data streams manually then integrate into new visualizations. Others highlighted the risk of costly errors in a manual process and said that the grouped and ranked results meant propagation was about ``sanity checking'' rather than making complex decisions. None of them had seen propagation-like features in other visualization platforms or tools, whilst those with industry experience suggested that the platforms they use, and their own working practice, could benefit from such features.}

\aar{
We also discussed the search and results user interface designs. When constructing queries, all used the quick keyword selection feature rather than type keywords; some said during the interview that this helped them create their search terms more quickly and meant they would not need to memorize keywords. The keyword colour coding in the search form seemed intuitive, although was most useful to participants when viewing the search results. Colour-coded keywords helped them decide which results to propagate visualization functions to, with users finding the keyword grouping to be especially useful; this meant they could scan and identify suitable groups, such that they could complete the tasks with less time and cognitive demand.
}

\section{Discussion}

This work was motivated by the significant need for visual analytics to support the emergency response to COVID-19. A significant volume and diversity of visual designs were required to support epidemiologists, modeling scientists and other domain experts in the SCRC, but we needed an approach that was feasible for a team of VIS volunteers in a context where timeliness was critical. As discussed in Section~\ref{sec:Overview}, we considered a number of solutions such as using existing visualization platforms, but the need for bespoke visualization and dashboard designs, and for strategically efficient use of volunteer resources, led us to the streamlined development and propagation approach outlined here.

\subsection{What did We Achieve?}
We observed, during our ongoing research project, a number of notable benefits of our method. We were able to: (i)~\textbf{re-purpose and reuse a given visualization} design in various contexts by propagating across numerous data streams, both for individual plots and composite dashboards, in an efficient yet controlled manner, thereby responding to the need for volume and diversity in visualizations; (ii)~ensure the suitability and efficacy of the visualizations offered through a \textbf{semi-automatic propagation process} that facilitates visualization quality assurance; (iii)~\textbf{streamline the visualization development} process by separating visualization development from infrastructure management; and (iv)~strategically \textbf{target our limited volunteer and development resources} to where they can make the most impact in a short time. We open up these points to further discussion in the following.

When visualizations are propagated across numerous different data streams or re-purposed within various dashboards, \textbf{quality assurance is of paramount importance}; visualizations need to be checked to ensure data streams (via their data types and keywords in our ontology) are suitable for the given plot and/or dashboard. This is especially important in the case of a visualization system developed to support the response to the pandemic, since the visualizations are involved in critical inference and decision-making scenarios. One can argue that automated visualization generation~\cite{Zhu:2020:VI} could be an alternative to streamline the visualization propagation process, but we have observed in our solution that a fully automated approach is not always reliable and can lead to unsuitable propagation. Given the importance of this task and such potential limitations in algorithmic methods, we developed a semi-automated approach that uses an ontology with algorithmic support for searching and ranking data streams for propagation, with a user interface that supports a 
infrastructure manager in assessing and approving the recommendations from the algorithms. 

\subsection{Importance of Roles and Separation of Concerns}
Throughout this project (and in our ongoing efforts), we were faced with a growing need for plots and dashboards for a range of data sources, while the resources for developing these were limited. Our approach addresses this in two ways. First, by limiting demand on VIS developer time by propagating a visual design to all datasets that will be useful to the domain experts. Second, by enabling a more strategic approach to resource management by \textbf{decoupling visualization design from data management.} This keeps visualization developers away from the complexities of the infrastructure and gives them more time to focus on designing and developing novel visualization capability.
Meanwhile, the data infrastructure and propagation workflows are managed by dedicated developers who are well-versed in the data streams and infrastructure, and are skilled at ensuring the quality of the propagation via quality assurance. These volunteers are not responsible for VIS design. This \textbf{separation of roles} is not only an effective use of developer time, but ensures the integrity of the final product. While we present these roles as distinct individuals in the paper, in reality, there may be overlaps and the same individual might be wearing multiple hats, e.g., a developer who is comfortable in designing and developing visualizations could also contribute to data management. 

\textbf{Our template-based propagation framework enables the separation of visualization design and implementation}. Following on from the growing trend of visualization specification languages~\cite{Satyanarayan:2017:TVCG}, this enables the design of plots to be explicitly specified and formulated without the constraints of the data infrastructure. This has a number of advantages, in that it is easier to ensure consistency across numerous plots and dashboards, and provides \textbf{consistency in how visualizations are presented over the web}, e.g., consistency in naming, titles, details of descriptions. Such an approach also makes the designs more transferable to contexts where new data streams and visual analytic needs may arise in short time. For instance, with the introduction of vaccination, it is possible to propagate several existing visualizations used for case/hospitalization numbers to this new context. 

\subsection{Upfront Costs}
Our approach to designing and developing a propagation mechanism did take significant time and resources to get the system in operation. The \textbf{implementation of the data and propagation infrastructure was a significant development effort} and, for a while, most of the development had to happen in the ``backstage'' with limited progress to demonstrate to domain experts in terms of visualization selection. However, our novel approach is a result of overcoming these technical challenges. Once the propagation system was functional, our approach was able to multiply the designs to various context and enable the rapid deployment a wide portfolio of visualizations. \textbf{While initial progress in terms of visualization offering would be slow}, our scalable and flexible approach ``future-proofs'' the system as new data products become available, to meet the varying VIS needs of domain experts.


\subsection{Generalizing Our Approach}
While our approach has been motivated by the ongoing pandemic, the proposed propagation approach and the workflow that our framework supports is transferable to different data-intensive settings, where there is demand for diverse and abundant visualization designs with large collections of data streams. The ontology-supported approach and underlying schema can be generalized and adapted to new data contexts. Our ontology models the data and VIS infrastructure, but \textbf{domain knowledge exists via data stream attributes} (i.e., keywords, descriptions), decoupled from the infrastructure implementation. Our system can rapidly transfer to a new domain by adding new data streams and capturing domain knowledge in their attributes, while a set of generic \textbf{visualizations (plots and dashboards) would be already available}. One potential benefit in such a ``transfer'' would be the ability to \textbf{propagate the visualization designs} and re-use them in suitable combinations within dashboards tailored for the specifics of the new application context. The transferability is also key for ensuring the preparedness of the visualization response in future situations where time-critical VIS is essential and could provide a solid foundation for further development.



 


\section{Conclusion}
This paper presents an ontology-based visualization development and propagation framework, with a streamlined workflow developed in response to the significant development challenges faced by the RAMPVIS volunteer visualization effort while responding to the COVID-19 pandemic. Our key challenge was to meet the need for a large number of diverse plots and dashboards, to meet the constantly evolving visual analytic requirements of domain experts in the Scottish COVID-19 Response Consortium. Meeting this challenge with scarce development and volunteer resources was only possible through a carefully designed infrastructure that streamlines the development process.

We do this through a visual design workflow that separates VIS development from the data infrastructure. Our ontology plays a key role in our infrastructure, capturing the relationships between data streams, VIS functions and web pages. We used an ontology-supported propagation process to allow a particular visualization to be rapidly deployed across numerous suitable data streams, instantly deploying them as interactive web pages. This enables a workflow that allows VIS volunteers focus their efforts on tasks they are most effective in. 

Our approach now enables the RAMPVIS consortium to offer a wide range of quality-assured plots and dashboards within a consistent presentation framework. With the changing demands of the ongoing pandemic management efforts (e.g., attention shifting to vaccination campaigns), our approach makes the visualization response from the consortium more resilient, responsive, and sustainable.  We are currently working closely together with SCRC to adapt our system to the rapidly changing nature of the pandemic and to improve our visualizations, dashboards, and user interfaces for use through the domain experts. In conclusion, we argue that our approach could serve as a blueprint for similar volunteer VIS efforts in future. In situations where the timely delivery of large-scale visualization is mission-critical, frameworks like these strengthen the key role that visualization plays in informing critical inference and decision-making.


%

\acknowledgments{\footnotesize{
This work was supported by EPSRC (EP/V054236/1). We would like to thank all volunteers from the SCRC and all VIS volunteers~\cite{RAMPVISvolunteers}. We would also like to thank Prof. N. W. John (U.~Chester) and Dr H. C. Purchase (U.~Glasgow) for their involvement in work of the generic support team. We are grateful to Dr R. Reeve (U.~Glasgow) and A. Brett (UKAEA) for their leadership in creating the SCRC data infrastructure that the VIS infrastructure depends on, and A. Lahiff and his STFC colleagues for maintaining the RAMP VIS VMs, and S. Michell (U.~Glasgow) for offering valuable advice on data products.}}

\bibliographystyle{abbrv-doi}

\bibliography{references}

\clearpage
\newpage
\noindent\huge%
\textsc{\textsf{\textbf{Supplementary Material}}}

\noindent\LARGE%
\textbf{Propagating Visual Designs to Numerous Plots and Dashboards}

\noindent\large%
~\\
Saiful Khan, University of Oxford, UK\\
Phong H. Nguyen, Redsift Ltd., UK\\
Alfie Abdul-Rahman, King's College London, UK\\
Benjamin Bach, Edinburgh University, UK\\
Min Chen, University of Oxford, UK\\
Euan Freeman, University of Glasgow, UK\\
Cagatay Turkay, University of Warwick, UK\\

\normalsize%

\appendix

\section{Algorithms for Computing Similarity Matrices}
\label{sec:algorithms-for-computing-similarity-matrices}

In this section we will describe the process of computing similarity matrices $\mat{S}_{rd}$ and $\mat{S}_{dd}$ defined in the paper and shown in \autoref{fig:illustration-grouping-ordering-datastream}. Matrix $\mat{S}_{rd}$ represents the pairwise similarity between the reference data streams and discovered data streams, and matrix $\mat{S}_{dd}$ represents pairwise similarities between each discovered data stream.
     
As described in the paper, each data stream (an OntoData instance) has several attributes: i.e., description, keywords, data type and API endpoint. Each attribute is a feature and we denote the number of features as $f$. The reference data streams (an ordered list) is denoted as a matrix $\mat{R} \in \mathbb{R}^{k \times f}$, where $k$ is the number of reference data streams. Discovered data streams are denoted as a matrix $\mat{D} \in \mathbb{R}^{n \times f}$, where $n$ is the number of discovered data streams. 

\subsection{Computing Matrix $\mat{S}_{rd}$}
\label{sec:computing-s_rd}

The ordering algorithm will use $\mat{S}_{rd}$ to sort the data streams within each group. To compute $\mat{S}_{rd}$, we first create a similarity matrix for each data stream feature and discovered data streams, then aggregate the four matrices.

\textbf{Data type similarity.}
We derive a feature vector $\vec{r}^{(t)} \in \mathbb{R}^{k}$, where $\vec{r}^{(t)}$ corresponds to the data type column of $\mat{R}$; $ \vec{r}^{(t)} = [ r^{(t)}_{1}, r^{(t)}_{2}, \dots r^{(t)}_{k} ]$. We derive another feature vector $\vec{d}^{(t)} \in \mathbb{R}^{n}$, where $\vec{d}^{(t)}$ corresponds to the data type column of $\mat{D}$; $\vec{d}^{(t)} = [d^{(t)}_{1}, d^{(t)}_{2}, \dots d^{(t)}_{n}]$.

A function $\phi$ computes the pairwise similarity matrix between the feature vectors $\vec{r}^{(t)}$ and $\vec{d}^{(t)}$ and is defined in \autoref{eq:data-type}.

\begin{equation} \label{eq:data-type}
\phi(\vec{r}^{(t)}, \vec{d}^{(t)}) \in \mathbb{R}^{k \times n} =  
    \begin{cases}
      1, & \text{if data types are similar}\\
      0, & \text{otherwise}
    \end{cases}
\end{equation}

\textbf{Keyword similarity.}
Similar to the data type vectors, we derive keyword feature vectors $\vec{r}^{(w)}$ and $\vec{d}^{(w)}$ which corresponds to the keyword column of matrix $\mat{R}$ and $\mat{D}$ respectively. The keywords attribute of a data stream contains a subset of a set of all keywords used to define data streams in the system. Therefore, Jaccard \cite{Jaccard:1912, Tanimoto:1958} similarity measurement function is used to compute the pairwise similarity matrix. A function $\psi$ computes the size of the intersection divided by the size of the union of two keywords sets, defined in \autoref{eq:keywords}.

\begin{equation}\label{eq:keywords}
    \psi(\vec{r}^{(w)}, \vec{d}^{(w)}) \in \mathbb{R}^{k \times n} = 
    \frac{ (\vec{r}^{(w)} \cup \vec{d}^{(w)}) } { ( \vec{r}^{(w)} \cap \vec{d}^{(w)}) }
\end{equation}

\textbf{Description similarity.} 
A description field is free-form text and can be represented as a collection of words or terms. Therefore, term frequency (tf) and inverse document frequency (idf) similarity measurement algorithms~\cite{Manning2008:Book} will be suitable here.


We derive feature vectors $\vec{r}^{(d)}$ and $\vec{d}^{(d)}$ which corresponds to the description column of $\mat{R}$ and $\mat{D}$ respectively.





Next, we derive a matrix $\mat{U}$, where each vector $\vec{u}_i \in \mat{U}$ is a $\mathrm{tf\mbox{-}idf}$ vector \cite{Manning2008:Book}  of $i$-th element of $\vec{r}^{(d)}$.
Similarly, we derive another matrix $\mat{V}$, where each vector $\vec{v}_j \in \mat{V}$ is a $\mathrm{tf\mbox{-}idf}$ vector of $j$-th element $\vec{d}^{(d)}$.

A function $\omega$ computes similarity by measuring Cosine similarity~\cite{Manning2008:Book} between $\mat{U}$ and  $\mat{V}$; defined in \autoref{eq:description}.

\begin{equation}\label{eq:description}
    \omega(\vec{r}^{(d)}, \vec{d}^{(d)}) \in \mathbb{R}^{k \times n}
    = \frac{  \mat{U} \mat{V} }{ \| \mat{U}\| \| \mat{V}\| }
\end{equation}

Given any two row vectors $\vec{u} \in \mat{U}$ and $\vec{v} \in \mat{V}$ the Cosine similarity between the vectors can be computed by \autoref{eq:cosine}.

\begin{equation}\label{eq:cosine}
    \frac{ \vec{u} \cdot \vec{v} }{ \| \vec{u} \| \| \vec{v} \| } 
    = \frac{ \sum_{x=1}^q u_x v_x }{ \sqrt{\sum_{x=1}^q u_x^2} \sqrt{\sum_{x=1}^q v_x^2} }
\end{equation}

\noindent where $u_x$ and $v_x$ are components of vector $\vec{u}$ and $\vec{v}$ respectively; and $q$ is the number of components (all possible words from the description fields).

\textbf{API endpoint similarity.}
We derive two feature vectors $\vec{r}^{(a)}$ and $\vec{d}^{(a)}$, which correspond to the API endpoint attribute columns of $\mat{R}$ and $\mat{D}$ respectively. RESTful API endpoints contain textual features such as a data stream server address, API route, and URL encoded parameters. An API endpoint can provide information about a data stream, such as its data product, component, source, and type (described in Section~\ref{sec:DataStreams}). We tokenize the endpoint fields to extract their terms (words) and use similar functions used for description field similarity measurement, $\omega({\vec{r}^{(a)}, \vec{d}^{(a)}}) \in \mathbb{R}^{k \times n}$, to compute a similarity matrix. 

\textbf{Aggregated similarity.}
We aggregate the four similarity matrices computed above. An aggregation function $\gamma$, defined in \autoref{eq:aggregate-rd}, computes the aggregated matrix, $\mat{S}_{rd} \in \mathbb{R}^{k \times n}$. This function takes a weighting average of the input matrices.

\begin{equation} \label{eq:aggregate-rd}
    \begin{split}
        \mat{S}_{rd} = \gamma(\mat{R}, \mat{D}) & = 
        [ \alpha \psi(\vec{r}^{(w)}, \vec{d}^{(w)}) + 
        \beta \omega(\vec{r}^{(d)}, \vec{d}^{(d)}) \\
         & \quad + \theta \omega(\vec{r}^{(a)}, \vec{d}^{(a)}) ]
         \odot \phi(\vec{r}^{(t)}, \vec{d}^{(t)})
    \end{split}
\end{equation}

\noindent where $\alpha$, $\beta$, and $\theta$ are scalar constants that define the relative weights of keywords, description, and endpoint fields in the similarity measurement, where $\alpha + \beta + \theta = 1 $. The pairwise similarity between any data type field is $0$ or $1$; therefore, for the keywords type we use the element-wise product (or Hadamard product), $\odot$, in the aggregation function.

\subsection{Computing Matrix $\mat{S}_{dd}$}
\label{sec:computing-s_dd}

The matrix $\mat{S}_{dd} \in \mathbb{R}^{n \times n}$ computes pairwise similarities between each discovered data streams. We use this matrix for creating uniform groups of similar data streams.

Computation of the matrix $\mat{S}_{dd}$ is almost similar to the computation steps applied for deriving the matrix $\mat{S}_{rd}$ in previous Section~\ref{sec:computing-s_rd}. 
For each feature, e.g., keyword, description, and API endpoint of the matrix $\mat{D}$, we derive three similarity matrices: $\psi(\vec{d}^{(w)}, \vec{d}^{(w)})$, $\omega(\vec{d}^{(d)}, \vec{d}^{(d)})$, and $\omega(\vec{d}^{(a)}, \vec{d}^{(a)})$ (using Equation~\ref{eq:keywords} and \ref{eq:description}).
Finally, the aggregation function, $\lambda(\mat{D}, \mat{D})$ aggregates the three matrices, defined in Equation~\ref{eq:aggregate-dd}.
A group can consist of data stream of different data types; therefore, we excluded the data type feature from the computation of $\mat{S}_{dd}$.

\begin{equation} \label{eq:aggregate-dd}
    \begin{split}
        \mat{S}_{dd} =  \lambda(\mat{D}, \mat{D}) & = \alpha \psi(\vec{d}^{(w)}, \vec{d}^{(w)}) + 
        \beta \omega(\vec{d}^{(d)}, \vec{d}^{(d)}) \\
        & \quad + \theta \omega(\vec{d}^{(a)}, \vec{d}^{(a)})
    \end{split}
\end{equation}

\section{Algorithms for Grouping and Ordering}
\label{app:algorithms-for-grouping-and-ordering}

Algorithm~\ref{algo:grouping-bruteforce} outlines the brute-force grouping approach described in Section~\ref{sec:Grouping}. Algorithm~\ref{algo:grouping-spectral} outlines the graph spectral grouping approach described in Section~\ref{sec:Grouping}. Algorithm~\ref{algo:ranking} outlines the procedure for sorting and ranking groups.

\begin{algorithm}[h!] \label{algo:grouping-bruteforce}
    \caption{A brute-force algorithm for grouping data streams}
    \SetAlgoLined
    \KwIn{A matrix $\mat{S}_{dd} \in \mathbb{R}^{n \times n}$; size of each group \sk{$k$}}
    \KwOut{Groups $\mat{G} \in \mathbb{R}^{k \times m}$}

    \BlankLine

    \DontPrintSemicolon
    \SetKwFunction{Group}{Group}
    \SetKwFunction{MaxIndices}{MaxIndices}
    \SetKwProg{Fn}{Function}{:}{}
    \SetKw{Continue}{continue}
    \SetCommentSty{mycommentfont}
    
    $n \leftarrow \| \mat{S}_{dd} \|$\; 

    \tcc{\sk{Number $m$ groups to construct}}
    \sk{$m = \left\lvert \frac{n}{k} \right\rvert$\;}
    
    $\mat{G} \leftarrow Null$\;
    $p \leftarrow 1$\;
    \BlankLine
    \For{$i \leftarrow 1$ \KwTo $n$}{
        $visited[i] \leftarrow False$\;
    }

    \BlankLine
    \For{$i \leftarrow 1$ \KwTo $n$}{
        \eIf{$visited[i]$ == $True$}{
            \Continue\;
        }{
            \tcc{Indices of $k$ max items of $i$-th row of $\mat{S}_{dd}$}
            $max\_k\_idx$ $\leftarrow$ {\MaxIndices{$\mat{S}_{dd}[i]$, $k$}}\;
            $\mat{G}[p] \leftarrow max\_k\_idx$
            
            $p \leftarrow p + 1$
            
            \BlankLine
            \tcc{Mark the grouped items as visited}
            \For{$j \leftarrow 1$ \KwTo $k$}{
                $idx \leftarrow max\_k\_idx[j]$\;
                $visited[idx] \leftarrow true$\;
            }
        }
    }
\end{algorithm}

\begin{algorithm}[h!]\label{algo:grouping-spectral}
    \caption{Spectral graph partitioning algorithm for grouping data streams}
    \SetAlgoLined
    \KwIn{A similarity matrix $\mat{S}_{dd} \in \mathbb{R}^{n \times n}$; size of each group \sk{$k$}}
    \KwOut{Groups $\mat{G} \in \mathbb{R}^{k \times m}$}

    \BlankLine
    
    \begin{itemize}[noitemsep]
        \item
        \sk{Number $m$ groups to construct, $m = \left\lvert \frac{n}{k} \right\rvert$.}
        \item
        Compute diagonal matrix $\mat{D}$ of $\mat{S}_{dd}$.
        \item 
        Compute Laplasian matrix, $\mat{L} = \mat{D} - \mat{S}_{dd}$.
        
        \item
        Given $m$ number of groups to create; compute the first $m$ eigenvectors $\vec{v}_1$, $\vec{v}_2$ \dots $\vec{v}_m$ of $\mat{L}$, such that $\mat{L}_v = \lambda \mat{D}_v$.
    
        \item
        Let $\mat{V} \in \mathbb{R}^{n \times m}$ be the matrix containing the vectors $\vec{v}_1$, $\vec{v}_2$, \dots $\vec{v}_m$ as columns.
        
        \item
        \sk{For $i = 1, 2, \dots n$, let $\vec{d_i} \in \mathbb{R}^{m}$ be the vector corresponding to the $i$-th row of $\mat{V}$.}
        
        \item
        Finally, use K-means algorithm to cluster $\vec{d_1}$, $\vec{d_2}$, \dots, $\vec{d_n}$ into $m$ groups $\vec{g_1}, \vec{g_2}, \dots \vec{g_m} \in \mat{G}$ and $\mat{G} \in \mathbb{R}^{k \times m}$.
    \end{itemize}
\end{algorithm}

\begin{algorithm}[h] \label{algo:ranking}
    \caption{Algorithm for sorting and ranking groups}
    \SetAlgoLined
    \KwIn{Groups $\mat{G} \in \mathbb{R}^{k \times m}$; matrix $\mat{S}_{rd} \in \mathbb{R}^{k \times n}$}
    \KwOut{Sorted and ranked $\mat{G}^{'} \in \mathbb{R}^{k \times m}$}
    \BlankLine

    \DontPrintSemicolon
    \SetKwFunction{Rank}{Rank}
    \SetKwProg{Fn}{Function}{:}{}
    \SetKw{Continue}{continue}
    \SetCommentSty{mycommentfont}
    \SetKwFunction{MaxIndices}{MaxIndices}
    \SetKwFunction{PriorityQAdd}{PriorityQAdd}
    
    $m \leftarrow \| \mat{G} \|$\;
    $k \leftarrow \| \mat{S}_{rd} \|$\;
    
    \tcc{Priority queue for queuing based on ranking}
    $\mat{G}^{'} \leftarrow Null$\; 
    
    \BlankLine
    \tcc{For each group}
    \For{$i \leftarrow 1$ \KwTo $m$}{
        $group \leftarrow \mat{G}[i]$\;
        $group\_sorted \leftarrow Null$\;
        $group\_score \leftarrow 0$
        
        \BlankLine
        \For{$j \leftarrow 1$ \KwTo $k$}{
            $col\_idx \leftarrow group[j]$\;
            $col\_vec \leftarrow \mat{S}_{rd}[:,col\_idx]$\;
            \tcc{The index of max element in $col\_vec$}
            $row\_idx \leftarrow$ {\MaxIndices{$col\_vec$, $1$}}\;
            $similarity \leftarrow col\_vec[row\_idx]$\;
            $group\_score \leftarrow group\_score$ + $similarity$\;
            $group\_sorted[row\_idx] \leftarrow col\_idx$\;
        }
    
        \BlankLine
        \tcc{Add the sorted group to priority queue}
        {\PriorityQAdd{$\mat{G}^{'}$, $group\_sorted$, $group\_score$}}\;
    }
\end{algorithm}
 
\section{Propagating Dashboards with Links}
\label{sec:workflow-for-searching-matching-links-for-dashboard}

In the paper, we described the process for propagating a visualization function with multiple data streams. Propagating a dashboard is more complicated, because of the need to match data streams (OntoData instances) \textit{and} web page links (OntoPage instances). If incorrectly matched, the visual designs in the dashboard would not be linked to the correct webpage.

\begin{figure}[h!] 
    \centering
    \includegraphics[width=1.0\columnwidth]{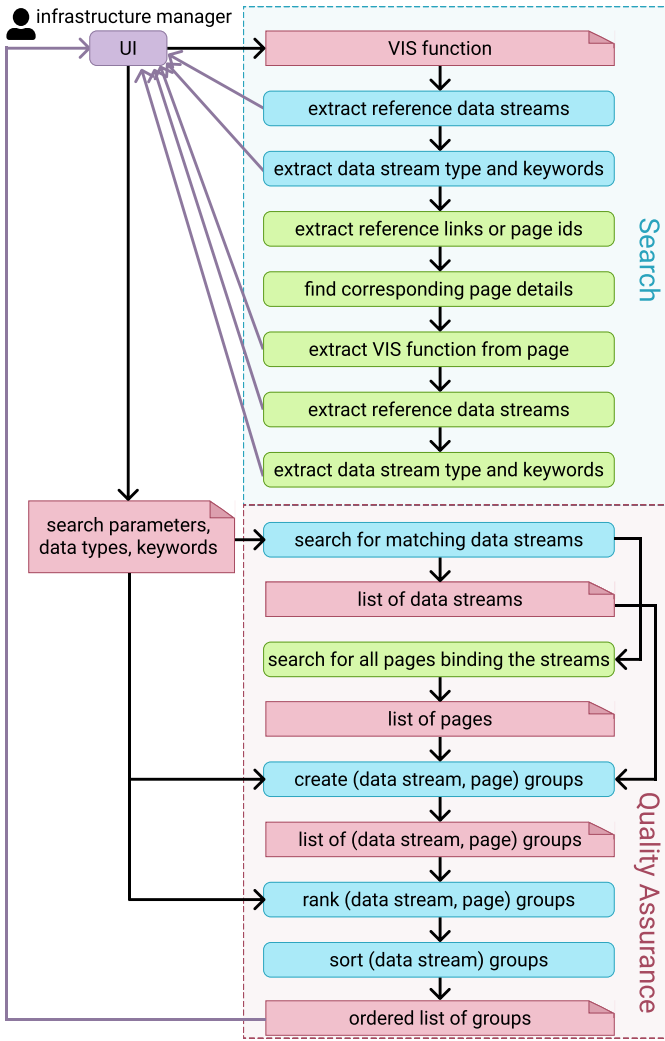}
    \caption{A flowchart illustrating a search and ranking workflow for finding relevant links and propagating the links to a dashboard.}
    \label{fig:flowchart-for-searching-matching-links-for-dashboard}
\end{figure}

\autoref{fig:flowchart-for-searching-matching-links-for-dashboard} shows the process used to propagate a dashboard with data streams and links. After the infrastructure manager selects a VIS function for propagation, its reference data streams and links are retrieved from the ontology. The metadata of the data streams and links are forwarded to the UI. While, the process of propagation for data streams is described in Section~\ref{sec:algorithm-support-for-vis-function-propagation}, the process of propagating links involves additional steps.

Links are web pages (i.e., OntoPage instances) and their attributes include a VIS function and data streams (described in Section~\ref{sec:visual-design-workflow}). Following the process described earlier in Section~\ref{sec:ui:search}, the infrastructure manager formulates a search query and discover a possible list of data streams. From the discovered data streams, a function (a)~creates possible groups of data streams and orders each group (as described in Section~\ref{sec:algorithm-support-for-vis-function-propagation} and Section~\ref{sec:algorithms-for-computing-similarity-matrices}).
For discovered data streams, another function scans the ontology to retrieve (b)~all possible pages or bindings visualizing the groups of (a). 
From the (a) and (b) list of possible groups, both data stream groups and page groups are created. 
We then use similar grouping, ordering, and ranking algorithms (described in Section~\ref{sec:algorithm-support-for-vis-function-propagation} and Section~\ref{sec:algorithms-for-computing-similarity-matrices}).

\end{document}